# A One-Parameter Diagnostic Classification Model with Familiar Measurement Properties

Matthew J. Madison, University of Georgia; mjmadison@uga.edu
Stefanie A Wind, University of Alabama; swind@ua.edu
Lientje Maas, Utrecht University; j.a.m.maas@uu.nl
Kazuhiro Yamaguchi, University of Tsukuba; kazuhir.ft@u.tsukuba.ac.jp
Sergio Haab, University of Georgia; sergio.haab@uga.edu

## Abstract

Diagnostic classification models (DCMs) are psychometric models designed to classify examinees according to their proficiency or non-proficiency of specified latent characteristics. These models are well-suited for providing diagnostic and actionable feedback to support formative assessment efforts. Several DCMs have been developed and applied in different settings. This study proposes a DCM with functional form similar to the 1-parameter logistic item response theory model. Using data from a large-scale mathematics education research study, we demonstrate that the proposed DCM has measurement properties akin to the Rasch and 1-parameter logistic item response theory models, including test score sufficiency, item-free and person-free measurement, and invariant item and person ordering. We discuss the implications and limitations of these developments, as well as directions for future research.

**A One-Parameter Diagnostic Classification Model with Familiar Measurement Properties**

Diagnostic classification models (DCMs; Rupp, Templin, & Henson, 2010), also known as cognitive diagnosis models (CDMs), are psychometric models designed to provide probabilistic classifications that indicate the proficiency status of examinees on specific latent traits, often termed *attributes*. In educational settings, these attribute proficiency classifications can be used to complement formative assessment efforts by highlighting students' strengths and areas to improve. Because of their criterion-referenced score interpretations and increased reliability and efficiency (Templin & Bradshaw, 2013), DCMs have been used in research studies and in operational assessment settings to support standards- and competency-based interpretations (Sessoms & Henson, 2020)

Similar to item response theory (IRT) models (e.g., Rasch, 1/2/3/4-parameter logistic (PL) models), there are several DCMs that make different assumptions about the item response generation process. DCMs applied in different settings may be selected for a variety of reasons, including substantive considerations, estimation complexity, ease-of-interpretation, and statistical fit. For example, the deterministic-input, noisy-and-gate model (DINA; Haertel, 1989; Junker & Sijtsma, 2001) is a commonly applied DCM that is often selected for its ease-of-interpretation and estimation simplicity. On the other hand, the log-linear cognitive diagnosis model (LCDM; Henson, Templin, & Willse, 2009) has been selected because of its generality and model refinement capabilities. When psychometric models are applied, there is a balance of statistical fit, stakeholder desires, and measurement properties that must be negotiated in selecting a model. We have seen this with assessment systems' choices of different IRT models; there are several IRT model-based operational assessment systems that use the Rasch model, although it is known that more complex IRT models will fit the data better. This decision is likely made to afford useful measurement properties such as test score sufficiency, a one-to-one and monotone test score - ability relationship, invariant measurement, and because the sacrifice in model fit may be negligible relative to the desired assessment results and interpretations (Andrich, 2002; Engelhard, 2013). In this context of psychometric model trade-offs and affordances, the purpose of this article is to present a DCM that sacrifices some flexibility and fit in order to obtain some of the aforementioned properties and demonstrate the utility of these properties with an empirical mathematics assessment.

      Before describing the proposed DCM, let us elaborate on the aforementioned properties and translate them into a DCM framework. One property of the Rasch model is that the total score is a sufficient statistic for examinee ability (Engelhard & Wang, 2020; Rasch, 1960). In a DCM framework, test score sufficiency would imply that the total test score is all that is required to obtain examinee probabilities of proficiency (and resulting classifications). If test score sufficiency is satisfied, it will follow that a test score threshold, or cutscore, for proficiency classifications can be determined. We note that in general, test score sufficiency does not hold for DCMs; proficiency classifications depend on the complete item response pattern (i.e., not only how many, but *which* items an examinee answers correctly). The Rasch model also has an invariant item ordering property, which states that if an item is more difficult than another item for any ability level, it must be more difficult for all ability levels (i.e., non-crossing item response curves). Relatedly, the Rasch model has an invariant person ordering property, which produces non-crossing person response functions. Practically, invariant person ordering means that the order of persons on the latent variable must be the same for all items. In a DCM framework, invariant item ordering would imply that if an item is more difficult than another item for any proficiency status, it must be more difficult for all the proficiency statuses.

Similarly, invariant person ordering would imply that an examinee with a higher proficiency status must have a higher correct response probability. Finally, the Rasch model maintains "item-free" and "sample-free" measurement of persons and items, respectively. In a DCM framework, item-free measurement would imply that classifications are not dependent on the specific set of items administered. And sample-free measurement would imply that DCM item parameter estimates are not dependent on the specific sample used in calibration. Using simulation studies, researchers have shown that DCMs provide item-free and sample-free estimates (de la Torre & Lee, 2010; Bradshaw & Madison, 2016). If all these properties are simultaneously attainable in a DCM framework, it could be beneficial for researchers and practitioners using DCMs to take advantage of them.

Now to the proposed DCM. As an ode to its elder sibling model (1-PL) and its foundation model (LCDM), we call this model the one-parameter LCDM (1-PLCDM). For items measuring a single attribute, the proposed 1-PLCDM estimates an intercept for each item (analogous to 1-PL difficulty parameters) and a single main effect across all items (analogous to the single 1-PL discrimination parameter). Readers familiar with foundation DCM developments may recognize this main effect constraint from the noisy-input, deterministic-and-gate (NIDA) (e.g., Junker & Sijtsma, 2001) and noisy-input, deterministic-or-gate (NIDO) (e.g., Templin, 2006) models. In the NIDA and NIDO models, item effects are modeled at the attribute level with equality constraints across items. The proposed 1-PLCDM is similar to the NIDA and NIDO models in that it constrains main effects across items measuring the same attribute, but unlike the NIDA and NIDO models, it freely estimates the item intercepts. We hypothesized that imposing this Rasch-like constraint might afford some of the Rasch model properties, and this was indeed the case. After describing the 1-PLCDM, we use data from a large-scale mathematics research study to demonstrate these properties. We conclude with a discussion of the results and implications for diagnostic modeling practice.

**Method**

In this section, we present the basic form of the proposed 1-PLCDM. Although it can be extended to multi-attribute cases, we present it in the single-attribute setting. We note here that the sufficiency property is guaranteed to hold in the single-attribute setting; it may hold in some multi-attribute settings, but not in general. We elaborate on this limitation in the discussion section.

*Proposed Model: 1-PLCDM*

The 1-PLCDM is a special case of the more general LCDM. In the single attribute case, the item response function for the LCDM is very similar to a unidimensional 2-PL item response function, except the latent trait is categorical. To see this similarity, Examinee $e$'s logit of a correct response to Item $i$ is given by:

$$logit(X_{ie} = 1) = \lambda_{i,0} + \lambda_{i,1}\alpha_e \tag{1}$$

In this equation, the intercept, $\lambda_{i,0}$, represents the log-odds of a correct response for examinees who are not proficient and the main effect, $\lambda_{i,1}$, represents the increase in log-odds of a correct response for examinees who are proficient relative to those who are not proficient. The main effect in Equation 1 is constrained to be larger than 0 to ensure that proficient examinees have a higher probability of correct response than non-proficient examinees. This parameter constraint also prevents class switching, a common issue in general mixture modeling (Lao, 2016; Redner and Walker, 1984). The person parameter in Equation 1 is $\alpha_e$ and represents the proficiency status, with $\alpha_e = 0$ representing non-proficiency and $\alpha_e = 1$ representing proficiency. In this way, the LCDM functions similarly to a reference-coded analysis of variance model, where the explanatory variable is the binary proficiency status.

To reduce to the 1-PLCDM, only a small adjustment needs to be made. In the LCDM, a main effect is estimated for every individual item, hence the main effect subscript *i* in Equation 1. In the 1-PLCDM, a single main effect is estimated across all items. Notice in Equation 2 below, the intercept is subscripted for individual items, but the main effect is not:

$$logit(X_i = 1) = \lambda_{i,0} + \lambda_1 \alpha_e \qquad (2)$$

In this way, the 1-PLCDM is similar to the 1-PL IRT model wherein a single discrimination parameter is estimated for all items.

## Empirical Demonstration

In this section, we use an empirical data set from a large-scale mathematics education research study to illustrate the presence of certain measurement properties in the 1-PLCDM. First, we describe the project from which the data originate, then we delve into the properties: test score sufficiency, item- and person-free measurement, and invariant person and item ordering.

### *Enhanced Anchored Instruction*

The data used in this study was collected in a mathematics education research study (Bottge, Ma, Gassaway, Toland, Butler, & Cho, 2014; Bottge, Toland, Gassaway, Butler, Choo, Griffen, & Ma, 2015). The study included 873 middle school students. The sample was mostly male (54%), mostly white (78%), and most students were in 7th grade (64%); 15% and 21% were in 6th and 8th grade, respectively. The overall goal of this project was to examine and evaluate an innovative instructional program called Enhanced Anchored Instruction (EAI; Bottge, Heinrichs, Chan, Mehta, & Watson, 2003) designed to improve students' problem-solving abilities. In EAI, students engage in authentic problem-solving sessions where they watch a 10 to 15-minute video that anchors the mathematical content in an interesting and relevant context. Students search the video for relevant information and use their mathematics knowledge to help the characters in the video develop a solution.

At the beginning and end of the instructional lessons, participating teachers administered several assessments including two standardized achievement tests and two researcher-developed tests. The assessment used in this empirical demonstration was a nine-item, researcher-developed, test designed specifically to measured students' problem-solving abilities. This test was open-ended and underwent piloting and several cycles of revisions in concert with mathematics teachers, researchers, and psychometricians. For this study, we used the pre-test responses, which were scored by the research team and had an interrater agreement of .93.

We used R software, specifically the *mirt* package (Chalmers, 2012) and the *mdirt* function (syntax available upon request), to estimate the 1-PLCDM and summarize results. The analyses below demonstrate the properties of the proposed model. We start with test score sufficiency and monotonicity.

*Test Score Sufficiency and Monotonicity*

Test score sufficiency is a property that states that the total test score (i.e., the sum score) contains all information required for estimation of the latent trait and is a well-known property of the Rasch model. Following from this property is that there is a one-to-one and monotonic relationship between test scores and Rasch ability estimates. Therefore, each examinee with the same test score will necessarily have the same Rasch ability estimate. Now we demonstrate that the 1-PLCDM also has the test score sufficiency property. In Figure 1, for the 1-PLCDM, each test score maps to exactly one posterior probability of proficiency (black squares). Additionally, the mapping is monotone: posterior probabilities are non-decreasing as total scores increase.

Using .50 as a probabilistic threshold for proficiency (the red horizontal line), the plot shows that in the 1-PLCDM, examinees with test scores greater than or equal to four received proficiency classifications. On the other hand, the test score-to-posterior probability mapping for the LCDM (grey circles) is not bijective; test scores between 1 and 8 map to multiple posterior probabilities of proficiency and examinees with total scores of two, three, four, and five had split classification results.

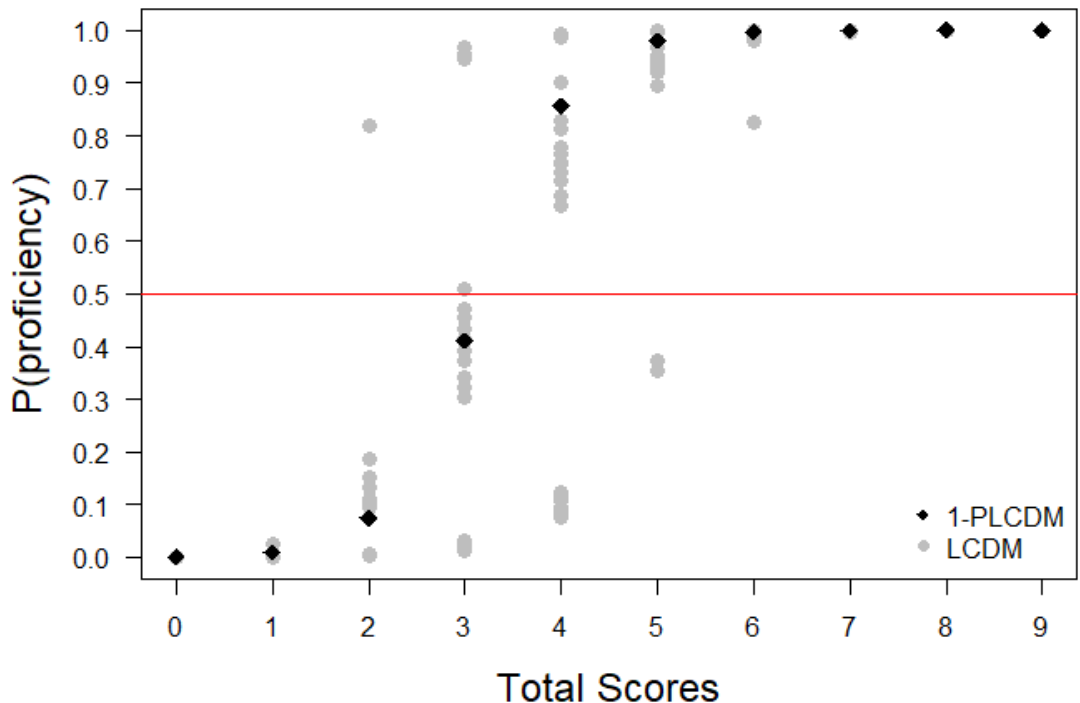

Figure 1. Total Score - Posterior Probability Scatter Plots

*Item- and Person-free Measurement*

To examine item-free measurement, we created two six-item tests from the original nine-item test. The Easy Test consisted of the six easiest items. The Hard Test consisted of the six hardest items. Difficulty was defined by the correct response probability for proficient examinees. We subset the responses to the both six-item tests, calibrated both tests, and compared the posterior probabilities and classifications from both tests. The proportions of proficient students for the Easy and Hard Tests were very similar at .50 and .53, respectively. The correlation between the tests' posterior probabilities of proficiency was .81 and the proficiency classification agreement between the two tests was .84. Even with the 33% reduction in test length from nine items to six items, we observed strong agreement in the person measurements from both tests. Therefore, these empirical results demonstrate the item-free measurement property of the 1-PLCDM.

To examine person-free measurement, we created two groups: one group of lower scoring examinees (less than median test score of three) and higher scoring examinees (higher than median test score of three). We randomly placed examinees with test scores equal to three in one of the low scoring or high scoring groups. Then we randomly selected ⅔ from the lower scoring group and ⅓ from the higher scoring group with replacement to form the Low Proficiency Group. Similarly, we selected ⅔ from the higher scoring group and ⅓ from the lower scoring group to form the High Proficiency Group. We then calibrated with both datasets. This process is similar to person-free calibration measurement investigations carried by Wright (1968) and de la Torre and Lee (2010).

Table 1 shows the 1-PLCDM item parameter point estimates and 95% confidence intervals from the complete dataset calibration and Low Proficiency and High Proficiency Group calibrations. For the intercepts, estimates from three calibrations were very similar, with the Complete Group estimates falling within both of the Low Proficiency or High Proficiency Group confidence intervals, and with the Low Proficiency and High Proficiency Group confidence intervals showing significant overlap. Similarly, for the lone main effect, the Complete Group estimate fell within the group confidence intervals, and the group confidence intervals overlapped. There was no apparent systematic bias in the parameter estimates (e.g., items appearing more difficult for Low Proficiency Group). We note that the demonstration of these invariance properties is not novel for DCMs; researchers have used simulation studies (de la Torre & Lee, 2010; Bradshaw & Madison, 2016) to show that DCMs have a theoretical invariance property and empirical demonstrations (e.g., de la Torre & Lee, 2010; Ravand, Baghaei, & Doebler, 2020) to show how real data with imperfect model fit can hinder the observation of the theoretical invariance property. We wanted, however, to illustrate that in addition to the other properties demonstrated, the 1-PLCDM also has the invariance properties that other DCMs possess. In this way, we show that although constrained, the 1-PLCDM maintains properties of general DCMs, and has added value with additional properties not possessed by other DCMs.

Table 1. Item Parameter Estimates for the Complete, Low Proficiency, and High Proficiency Samples

| Item | Complete $\lambda_0$ | Low Proficiency $\lambda_{0(.025)}$ | $\lambda_{0(.975)}$ | High Proficiency $\lambda_{0(.025)}$ | $\lambda_{0(.975)}$ | Complete $\lambda_1$ | Low Proficiency $\lambda_{1(.025)}$ | $\lambda_{1(.975)}$ | High Proficiency $\lambda_{1(.025)}$ | $\lambda_{1(.975)}$ |
|---|---|---|---|---|---|---|---|---|---|---|
| 1 | -0.92 | -1.13 | -0.79 | -1.21 | -0.67 | | | | | |
| 2 | -2.23 | -2.47 | -2.05 | -2.52 | -1.99 | | | | | |
| 3 | -1.13 | -1.44 | -1.08 | -1.49 | -0.95 | | | | | |
| 4 | -0.81 | -1.02 | -0.68 | -1.13 | -0.60 | 2.15 | 2.11 | 2.39 | 2.01 | 2.36 |
| 5 | -4.87 | -5.39 | -4.48 | -4.97 | -4.25 | | | | | |
| 6 | -0.21 | -0.44 | -0.11 | -0.46 | 0.05 | | | | | |
| 7 | -2.05 | -2.36 | -1.95 | -2.40 | -1.86 | | | | | |
| 8 | -2.40 | -2.56 | -2.14 | -2.86 | -2.33 | | | | | |

*Note*. $\lambda_0$ is the intercept; $\lambda_1$ is the main effect. There is only one main effect ($\lambda_1$) in the 1-PLCDM. Value in parentheses is the confidence limit for the 95% confidence interval.

### *Invariant Person and Item Ordering*

Invariant person ordering is a property that states that examinees are ordered in the same way with respect to response probabilities across all items. In other words, if Examinee A has a higher response probability than Examinee B on any single item, this must be the case for all items (Wright & Masters, 1982). This property manifests itself in non-crossing person response functions (Engelhard, 2013). In a DCM framework, this property follows immediately from the main effects being constrained greater than zero. This constraint forces each successive proficiency level to have a higher correct response probability than the previous proficiency level. We move to the more interesting property: invariant item ordering.

Invariant item ordering is a property that states the items are ordered in the same way with respect to correct response probabilities for all examinees. In other words, if Item $I$ is more difficult than Item $J$ for any single examinee, this must be the case for all examinees. For the Rasch and 1-PL models, this property manifests itself in non-crossing item response curves (Engelhard, 2013; Wright & Masters, 1982). For the 1-PLCDM, and all DCMs more generally, there is no item response curve because attributes are categorical. Nevertheless, we can demonstrate this property visually. Figure 2 displays the nine item characteristic bar charts for the 1-PLCDM sorted by the probability correct for non-proficient examinees.

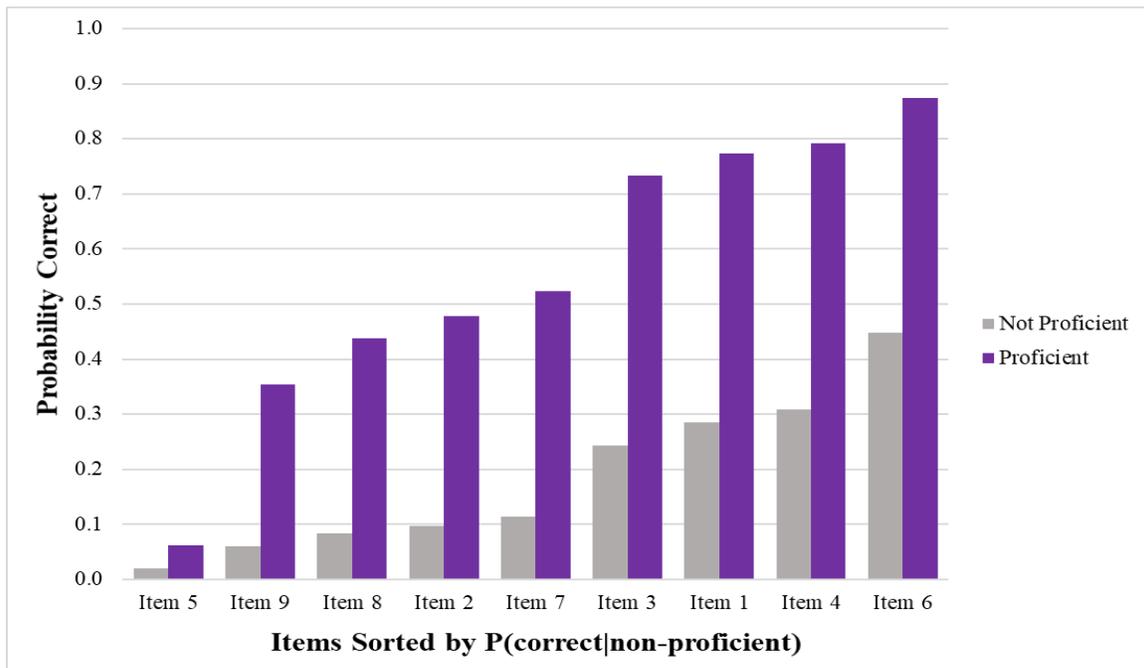

**Figure 2. 1-PLCDM Item Characteristic Bar Chart**

Notice that the items are also sorted by the probability correct for proficient examinees. This indicates the items are ordered in the same way for all examinee proficiency levels. On the other hand, Figure 3 displays the nine item characteristic bar charts for the LCDM sorted by the probability correct for non-proficient examinees. In Figure 3, notice the same property does not hold for the LCDM. For the LCDM, Item 6 is harder than Item 1 for non-proficient examinees, but Item 6 is easier than Item 1 for proficient examinees. Therefore, invariant item ordering is violated.

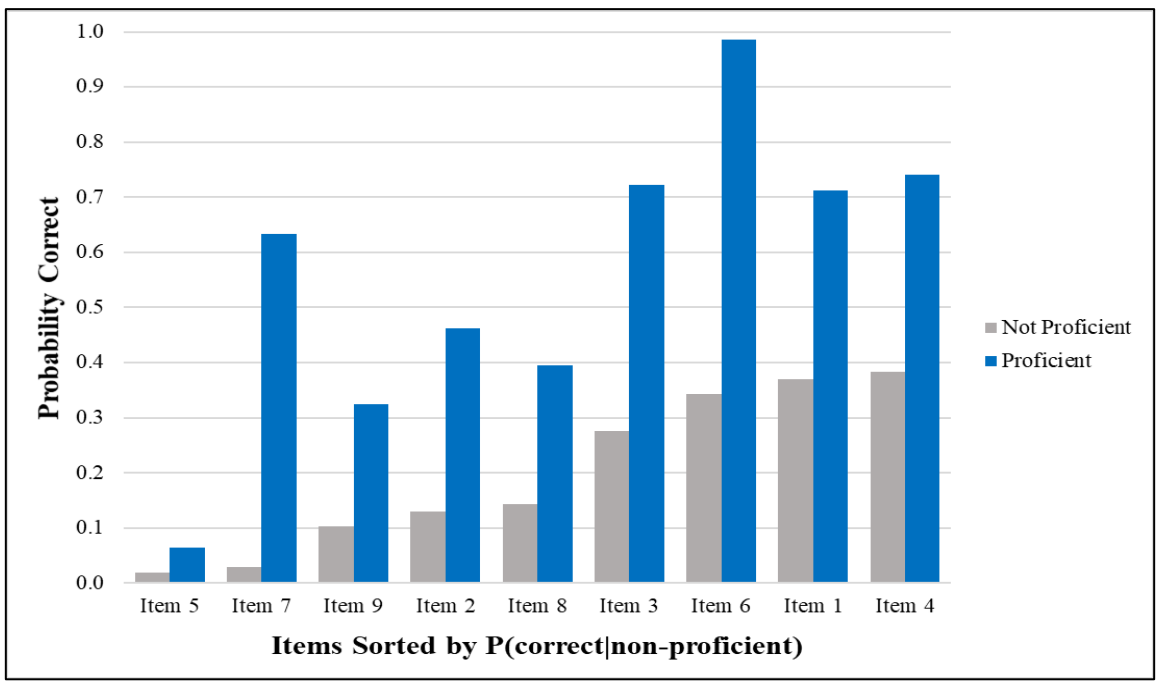

**Figure 3. LCDM Item Characteristic Bar Chart**

*Summary of 1-PLCDM Measurement Properties*

In this empirical demonstration, we have illustrated that the proposed 1-PLCDM has measurement properties most often associated with the Rasch and 1-PL models: test score sufficiency, item- and person-free measurement, and invariant item and person ordering. We demonstrated that the total test score was sufficient for estimating posterior probabilities of proficiency and that the test score - posterior probability mapping was bijective, thereby producing a model-estimated test score threshold for proficiency. We demonstrated that item parameters were invariant across samples with different proficiency levels and that examinee classifications were invariant across tests with different difficulty levels. Lastly, we demonstrated that the 1-PLCDM has invariant item and person ordering. To our knowledge, the 1-PLCDM is the only DCM to simultaneously possess all of these properties. These properties are particularly useful in operational contexts, where parsimony, interpretation and articulation, and invariance are paramount.

## Discussion

This study introduced a new DCM: the 1-parameter log-linear cognitive diagnosis model (1-PLCDM). Put simply, the 1-PLCDM is the categorical latent trait analogue of the 1-PL IRT model; it estimates a single main effect across all items, analogous to the single discrimination parameter estimated by the 1-PL IRT model. Using a researcher-developed mathematics assessment for middle school students, we illustrated that the 1-PLCDM has measurement properties most commonly associated with the Rasch and 1-PL IRT models. Namely, we demonstrated that the 1-PLCDM has a test score sufficiency property, which produces a one-to-one and monotonic relationship between raw total scores and posterior probabilities of proficiency, as well as a cutscore for proficiency. The 1-PLCDM exhibited item- and person-free measurement of persons and items, which implies that examinee classifications are the same regardless of the item difficulties and item parameter estimates are the same regardless of the sample characteristics. Finally, we showed that 1-PLCDM possesses item invariant person ordering and person invariant item ordering.

This study can be regarded as a first step in learning more about the 1-PLCDM, particularly as it relates to generalization of the model. In our explorations, the test score sufficiency property was not upheld in all cases when the assessment had multiple attributes or complex items. We hypothesize that in these cases, attribute correlations are the culprit; the posterior probability of proficiency for one attribute is impacted by its relationship with other attributes being modeled. This is a significant limitation because many DCM applications model multiple attributes. Furthermore, one of the main benefits of DCMs is how easily they accommodate multiple attributes and complex items measuring multiple attributes. Future work could examine the conditions under which these properties can be generalized. Of course, mathematical derivations and proofs (or refutations) of these properties would be useful, as well.

    Another issue to be investigated is the performance and relative robustness of the 1-PLCDM when its assumptions are violated. Specifically, it would be interesting to assess classification accuracy and reliability when the main effect constraint is violated to various degrees. Studies have shown that some DCMs can provide accurate and reliable classifications, even in the presence of certain model misspecifications (Kunina-Habenicht, Rupp, & Wilhelm, 2012; Madison & Bradshaw, 2018). If the 1-PLCDM can be shown to be robust, it would be a positive indicator for its practical application and make it appealing for stakeholders who want to apply it in various assessment settings. While more work, both empirical and theoretical, is required to fully realize the exciting potential of this model development, we hope that this study provides some initial insights into its application.